\documentclass[conference]{IEEEtran}

\IEEEoverridecommandlockouts

\usepackage{amsmath,amssymb}
\usepackage{graphicx}
\usepackage{booktabs}
\usepackage{enumitem}
\usepackage{cite}
\usepackage{url}
\usepackage{hyperref}
\makeatletter
\def\section{\@startsection{section}{1}{\z@}%
{1.1ex plus 0.8ex minus 0.4ex}%
{0.6ex plus 0.3ex minus 0.2ex}%
{\normalfont\normalsize\centering\scshape}}%
\def\subsection{\@startsection{subsection}{2}{\z@}%
{1.1ex plus 0.8ex minus 0.4ex}%
{0.6ex plus 0.3ex minus 0.2ex}%
{\normalfont\normalsize\itshape}}%
\def\subsubsection{\@startsection{subsubsection}{3}{\parindent}%
{0.2ex plus 0.1ex minus 0.1ex}%
{0.1ex}%
{\normalfont\normalsize\itshape}}%
\makeatother

\setlist[itemize]{leftmargin=*,noitemsep,topsep=2pt}
\setlist[enumerate]{leftmargin=*,noitemsep,topsep=2pt}
\begin{document}

\title{Occlusion-Aware Multimodal Beam Prediction and Pose Estimation for mmWave V2I}
\author{\IEEEauthorblockN{
Abidemi Orimogunje\IEEEauthorrefmark{1}\IEEEauthorrefmark{2},
Hyunwoo Park\IEEEauthorrefmark{2},
Kyeong-Ju Cha\IEEEauthorrefmark{2},
Igbafe Orikumhi\IEEEauthorrefmark{2},
Sunwoo Kim\IEEEauthorrefmark{2},
Dejan Vukobratovic\IEEEauthorrefmark{3}
}
\IEEEauthorblockA{\IEEEauthorrefmark{1}
African Center of Excellence in Internet of Things, University of Rwanda, Rwanda}
\IEEEauthorblockA{\IEEEauthorrefmark{2}
Department of Electronic Engineering, Hanyang University, South Korea}
\IEEEauthorblockA{\IEEEauthorrefmark{3}
Faculty of Technical Sciences, University of Novi Sad, Serbia.}
}
\maketitle
\begin{abstract}
We propose an occlusion-aware multimodal learning framework that is inspired by simultaneous localization and mapping (SLAM) concepts for trajectory interpretation and pose prediction. Targeting mmWave vehicle-to-infrastructure (V2I) beam management under dynamic blockage, our Transformer-based fusion network ingests synchronized RGB images, LiDAR point clouds, radar range--angle maps, GNSS, and short-term mmWave power history. It jointly predicts the receive beam index, blockage probability, and 2D position using labels automatically derived from 64-beam sweep power vectors, while an offline LiDAR map enables SLAM-style trajectory visualization. On the 60\,GHz DeepSense~6G Scenario~31 dataset, the model achieves 50.92\% Top-1 and 86.50\% Top-3 beam accuracy with 0.018\,bits/s/Hz spectral-efficiency loss, 63.35\% blocked-class F1, and 1.33\,m position RMSE. Multimodal fusion outperforms radio-only and strong camera-only baselines, showing the value of coupling perception and communication for future 6G V2I systems.
\end{abstract}
\section{Introduction}

Autonomous vehicles (AVs) in dense urban environments need accurate localization and reliable high-rate wireless links. These capabilities enable map updates, cooperative perception, and cloud-assisted planning~\cite{rs15041156}. Simultaneous localization and mapping (SLAM) is therefore central to AV perception, and light detection and ranging (LiDAR), camera, and radar-based methods achieve high accuracy in many scenarios~\cite{10742953,10287946}. Yet most SLAM pipelines remain communication-agnostic. Sensing and communication are still engineered as separate subsystems, even though future 5G/6G vehicle-to-infrastructure (V2I) deployments will increasingly share hardware, computation, and scene understanding across both stacks~\cite{10620740,9662195}.

This separation is particularly limiting for mmWave V2I. mmWave links can suffer abrupt throughput drops when the line-of-sight (LoS) path is blocked by vehicles, pedestrians, or roadside structures~\cite{11174444}. Conventional beam training that relies mainly on radio feedback can be slow and brittle under fast dynamics and intermittent blockage~\cite{10757880}. Unimodal beam prediction can also fail in occluded or visually ambiguous scenes~\cite{10816687}. These challenges motivate an \emph{occlusion-aware multimodal} approach, where a shared latent state fuses multi-sensor perception with short-term radio context to support both pose estimation and beam management.

Prior work addresses parts of this objective but rarely the full problem. Radio-only localization and beam-aware schemes exploit angle-of-arrival (AoA)~\cite{orimogunje2025mobilityawarelocalizationmmwavechannel,10561505}, angle-difference-of-arrival (ADoA)~\cite{10880651}, and virtual anchors~\cite{7925882,8057183}. These methods can be accurate in controlled settings, but they often assume static layouts and rely exclusively on radio observations. Semantic radio SLAM extends this line by estimating pose and labeling interaction points as reflectors or scatterers~\cite{10880651,11251132}, but it typically does not integrate complementary non-radio sensors for beam selection and occlusion handling. In contrast, multi-sensor SLAM stacks for autonomous driving fuse LiDAR, cameras, radar, and global navigation satellite system (GNSS) signals for robust localization under partial occlusion~\cite{11133431,9939167}. However, they rarely incorporate mmWave measurements or communication-centric objectives such as beam alignment and spectral-efficiency loss.

This paper bridges these directions using the real 60\,GHz DeepSense~6G Scenario~31 V2I dataset~\cite{10144504}. The dataset provides synchronized mmWave beam-sweep power vectors, RGB images, LiDAR point clouds, frequency-modulated continuous-wave (FMCW) radar features, and position readings along an urban street. We design a Transformer-based multimodal fusion network that processes per-snapshot observations and learns a shared latent representation. From this representation, it jointly predicts the receive beam index, blockage probability, and 2D vehicle pose. Beam and blockage labels are derived automatically from measured 64-beam power vectors. An offline LiDAR map is further used to visualize predicted trajectories in a SLAM-style manner. The main contributions are:
\begin{itemize}
  \item We cast receive-beam prediction, blockage detection, and 2D pose estimation as a unified multi-task learning problem over a shared latent state built from camera, LiDAR, radar, GNSS, and short-term mmWave power history.
  \item We develop a Transformer-based fusion architecture with modality-specific encoders and fully automatic beam/blockage label generation from beam-sweep measurements, yielding an occlusion-aware representation for mmWave V2I.
  \item We evaluate on DeepSense Scenario~31 against radio-only and single-sensor baselines, showing improved beam alignment, reduced spectral-efficiency loss, and stronger blockage and pose performance under multimodal fusion.
\end{itemize}
\section{System Model and Problem Formulation}
\label{sec:system_model}
We consider a mmWave V2I link consistent with DeepSense~6G Scenario~31~\cite{10144504}, which consists of a stationary infrastructure unit with synchronized multi-modal sensing and mmWave array receiver and a moving vehicle unit with transmitting phase array see Fig. 1. The mmWave receiver employs a phased array and (during data collection) sweeps a fixed receive-beam codebook of size $B=64$ at each time index $t\in\{1,\dots,T\}$ to measure the received power across beams. In addition, synchronized RGB, LiDAR, radar, and position measurements are available. Our learning objective is to \emph{predict} the optimal beam, blockage state, and vehicle pose without requiring an exhaustive beam sweep at inference time.
\begin{figure}
    \centering
\includegraphics[width=0.9\linewidth]{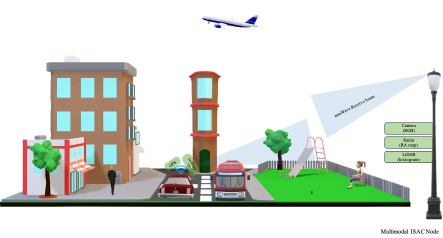}
 \caption{Illustration of the mmWave V2I scenario and multimodal integrated and sensing (ISAC) node used for occlusion-aware receive-beam selection and 2D vehicle localization.}
    \label{fig:systemmodel}
\end{figure}
\subsection{State and Map Representation}
We adopt a base station (BS)-centric local East-North frame and focus on planar motion. The vehicle 2D position at time $t$ is
\begin{equation}
  \mathbf{s}_t =
  \begin{bmatrix}
    x_t \\[2pt] y_t
  \end{bmatrix}
  \in \mathbb{R}^2.
\end{equation}
A GNSS receiver provides a noisy position measurement
\begin{equation}
  \tilde{\mathbf{s}}_t = \mathbf{s}_t + \mathbf{n}_t^{\mathrm{gps}},
\end{equation}
where $\mathbf{n}_t^{\mathrm{gps}}\in\mathbb{R}^2$ captures residual GNSS errors after projection to the local frame.
The static environment is represented by a LiDAR-based map
\begin{equation}
  \mathcal{M}_L = \{\boldsymbol{\ell}_i\}_{i=1}^{N_L}, \quad
  \boldsymbol{\ell}_i \in \mathbb{R}^3,
\end{equation}
constructed offline by aggregating and aligning LiDAR scans over the drive. Dynamic objects are not explicitly stored in $\mathcal{M}_L$; their impact on occlusion and multipath is handled implicitly by the learned model.
\subsection{Sensor and Observation Model}
At each time $t$, the system collects:
\begin{itemize}[leftmargin=*]
  \item a mmWave received-power vector $\mathbf{r}_t \in \mathbb{R}^B$, where $r_t(b)$ is the received power on beam $b$;
  \item an RGB image $I_t$;
  \item a LiDAR point cloud $L_t = \{\boldsymbol{\ell}_{t,i}\}_{i=1}^{N_t}$;
  \item an FMCW radar measurement $R_t$, processed into a range--angle magnitude map $\tilde{R}_t$;
  \item a GNSS reading, projected to the local 2D measurement $\tilde{\mathbf{s}}_t$.
\end{itemize}
For prediction, we form the model input as;
\begin{equation}
  \mathbf{o}_t =
  \big(
    I_t,\,
    L_t,\,
    \tilde{R}_t,\,
    \tilde{\mathbf{s}}_t,\,
    \mathbf{r}_{t-1}
  \big),
  \label{eq:obs_tuple}
\end{equation}
where $\mathbf{r}_{t-1}$ provides short-term radio context (set to $\mathbf{0}$ for $t=1$). The current sweep vector $\mathbf{r}_t$ is used \emph{only} to construct supervision labels and evaluation metrics.
\subsection{Beam and Blockage Labelling}
Using the beam sweep at time $t$, the oracle receive beam index is defined as
\begin{equation}
  b_t^\star = \arg\max_{b\in\{1,\dots,B\}} r_t(b),
\end{equation}
which provides the ground-truth label for beam prediction.
To generate blockage labels, we define the maximum received power
\begin{equation}
  P_t^{\max} = \max_{b} r_t(b).
\end{equation}
Let $\tau$ be a threshold chosen as a fixed percentile (e.g., 20th) of $\{P_t^{\max}\}$ over the training set. The binary blockage label is
\begin{equation}
  y_t^{\mathrm{blk}} =
  \begin{cases}
    1, & P_t^{\max} < \tau \quad \text{(blocked / low power)} \\
    0, & P_t^{\max} \ge \tau.
  \end{cases}
\end{equation}
For performance evaluation, the power on beam $b$ is mapped to an approximate spectral efficiency (SE),
\begin{equation}
  \mathrm{SE}_t(b) = \log_2\!\big(1 + \mathrm{SNR}_t(b)\big),
\end{equation}
where $\mathrm{SNR}_t(b)$ is computed from $r_t(b)$ under a fixed noise power (after converting to linear scale if the stored power values are in dB). The spectral-efficiency loss incurred by predicting beam index $\hat{b}_t$ instead of $b_t^\star$ is
\begin{equation}
  \Delta\mathrm{SE}_t
  = \mathrm{SE}_t(b_t^\star) - \mathrm{SE}_t(\hat{b}_t),
  \label{eq:se_drop_def}
\end{equation}
and we report the test-set average of $\Delta\mathrm{SE}_t$.
\subsection{Learning Objective}
We learn a snapshot-based mapping from multimodal observations to communication and localization outputs. Let $f_\theta$ denote the fusion network with parameters $\theta$. For each $\mathbf{o}_t$ in~\eqref{eq:obs_tuple}, the network produces
\begin{equation}
  \big(
    \mathbf{z}_t,\,
    v_t,\,
    \hat{\mathbf{s}}_t
  \big)
  = f_\theta(\mathbf{o}_t),
\end{equation}
where $\mathbf{z}_t \in \mathbb{R}^B$ are beam logits, $v_t\in\mathbb{R}$ is a blockage logit, and $\hat{\mathbf{s}}_t\in\mathbb{R}^2$ is the estimated 2D position.
Beam and blockage probabilities are
\begin{align}
  \boldsymbol{\pi}_t &= \mathrm{softmax}(\mathbf{z}_t), \\
  q_t                &= \sigma(v_t),
\end{align}
with $\boldsymbol{\pi}_t \in \mathbb{R}^B$ and $q_t\in[0,1]$. The predicted beam and hard blockage decision are
\begin{align}
  \hat{b}_t       &= \arg\max_b \boldsymbol{\pi}_t(b), \\
  \hat{y}_t^{\mathrm{blk}} &= \mathbb{1}[q_t \ge 0.5].
\end{align}
Given a supervised dataset of inputs, beam labels, blockage labels and reference positions,
$\mathcal{D} = \{(\mathbf{o}_t, b_t^\star, y_t^{\mathrm{blk}}, \mathbf{s}_t)\}_{t=1}^T$,
we train $f_\theta$ with a multi-task loss:
\begin{equation}
  \mathcal{L}(\theta)
  = \lambda_{\text{beam}} \mathcal{L}_{\text{beam}}(\theta)
  + \lambda_{\text{blk}}  \mathcal{L}_{\text{blk}}(\theta)
  + \lambda_{\text{pose}} \mathcal{L}_{\text{pose}}(\theta),
  \label{eq:learn}
\end{equation}
where $\mathcal{L}_{\text{beam}}$ is cross-entropy between $\boldsymbol{\pi}_t$ (beam classification loss) and $b_t^\star$, $\mathcal{L}_{\text{blk}}$ is binary cross-entropy between $q_t$ and $y_t^{\mathrm{blk}}$ (blockage classification loss) and $\mathcal{L}_{\text{pose}}$ is mean-squared error between $\hat{\mathbf{s}}_t$ and $\mathbf{s}_t$ (pose regression loss). The weights $\lambda_{\text{beam}},\lambda_{\text{blk}},\lambda_{\text{pose}}\ge 0$ control the trade-off across tasks.
The 2D localization root mean square error (RMSE) reported in Section~\ref{sec:results} is
\begin{equation}
\scalebox{0.9}{$
  \mathrm{RMSE}
  = \sqrt{\frac{1}{T}
    \sum_{t=1}^T \big\|\hat{\mathbf{s}}_t - \mathbf{s}_t\big\|_2^2}
$}
\label{eq:pose_rmse_def}
\end{equation}
computed on the held-out test set.
\section{Proposed Method}
\label{sec:method}
This section instantiates the learning formulation of Section~\ref{sec:system_model} on DeepSense~6G Scenario~31. We describe how each modality is preprocessed and embedded into a $d$-dimensional token, then present the Transformer fusion module and task heads that implement $f_\theta$, and finally outline training and LiDAR-based trajectory visualization.
\subsection{Multimodal Tokens: Preprocessing and Encoders}
For each synchronized time index $t$, the raw measurements in Section~\ref{sec:system_model} are converted into the observation tuple $\mathbf{o}_t$ in~\eqref{eq:obs_tuple}. Each modality is lightly preprocessed and encoded into a $d$-dimensional feature vector; all encoders are trained jointly with the fusion module.

\textbf{Camera (RGB):} The RGB image $I_t$ is resized to $224\times224$, converted to a tensor, and normalized using ImageNet statistics. A ResNet-18 backbone (pre-trained on ImageNet) extracts a global average-pooled feature, which is projected via a linear layer to $\mathbf{h}_t^{\text{img}} \in \mathbb{R}^d$.

\textbf{LiDAR:} The point cloud $L_t$ is filtered to remove invalid points and randomly subsampled to a fixed number of 3D points for efficiency. A PointNet-style encoder applies shared layers with ReLU to each point and aggregates using global max-pooling, yielding $\mathbf{h}_t^{\text{lidar}} \in \mathbb{R}^d$.

\textbf{Radar:} The FMCW radar cube $R_t$ is converted to a 2D range--angle magnitude map $\tilde{R}_t$ by applying FFTs over fast-time and antenna channels, followed by magnitude and normalization. A lightweight CNN with strided convolutions and global average pooling maps $\tilde{R}_t$ to $\mathbf{h}_t^{\text{radar}} \in \mathbb{R}^d$.

\textbf{GNSS:} Latitude/longitude are projected to the BS-centred local frame to form the 2D input $\tilde{\mathbf{s}}_t$. A two-layer MLP with batch normalization encodes $\tilde{\mathbf{s}}_t$ into $\mathbf{h}_t^{\text{gps}} \in \mathbb{R}^d$.

\textbf{mmWave power history:} During data collection, an exhaustive 64-beam sweep yields the received-power vector $\mathbf{r}_t \in \mathbb{R}^B$ (with missing entries set to zero if needed), which is used only to derive $b_t^\star$ and $y_t^{\mathrm{blk}}$ (Section~\ref{sec:system_model}). To mimic predictive operation, the network input includes only the previous vector $\mathbf{r}_{t-1}$ (or $\mathbf{0}$ for $t=1$). A two-layer MLP encodes $\mathbf{r}_{t-1}$ into $\mathbf{h}_t^{\text{mmw}} \in \mathbb{R}^d$, capturing short-term radio context and implicit occlusion cues.
These encoders also define the unimodal baselines by retaining a single encoder and replacing the fusion module with a shallow prediction head.
\subsection{Transformer-Based Fusion and Task Heads}
The five modality features at time $t$ are stacked into a token sequence
\begin{equation}
  \mathbf{H}_t =
  \big[
    \mathbf{h}_t^{\text{img}},
    \mathbf{h}_t^{\text{lidar}},
    \mathbf{h}_t^{\text{radar}},
    \mathbf{h}_t^{\text{gps}},
    \mathbf{h}_t^{\text{mmw}}
  \big] \in \mathbb{R}^{5 \times d}.
\end{equation}
A learnable classification token $\mathbf{h}^{\text{cls}}\in\mathbb{R}^d$ is prepended, and a token-type embedding encodes the modality identity (camera, LiDAR, radar, GNSS, mmWave, CLS). The resulting sequence is processed by a Transformer encoder with multi-head self-attention and feed-forward layers.
The output at the CLS position, $\mathbf{h}_t^{\text{CLS}} \in \mathbb{R}^d$, serves as a compact multimodal state summarizing geometry, occlusion cues, and radio context. Three linear heads then produce the beam logits, blockage logit, and pose estimate:
\begin{align}
  \mathbf{z}_t      &= W_{\text{beam}} \mathbf{h}_t^{\text{CLS}} + \mathbf{b}_{\text{beam}}, \\
  v_t              &= \mathbf{w}_{\text{blk}}^\top \mathbf{h}_t^{\text{CLS}} + b_{\text{blk}}, \\
  \hat{\mathbf{s}}_t &= W_{\text{pose}} \mathbf{h}_t^{\text{CLS}} + \mathbf{b}_{\text{pose}}.
\end{align}
Softmax and sigmoid activations yield the probabilities used in $\mathcal{L}_{\text{beam}}$ and $\mathcal{L}_{\text{blk}}$.
\subsection{Training Strategy}
All parameters $\theta$ are optimized with the multi-task loss $\mathcal{L}(\theta)$ in~\eqref{eq:learn}. We use AdamW with weight decay and gradient clipping. The dataset is split by sequence into training, validation, and test sets to reduce temporal leakage; within each sequence, the temporal ordering is preserved so that $\mathbf{r}_{t-1}$ corresponds to the immediately preceding snapshot.
Because blocked snapshots are relatively scarce, the positive class in $\mathcal{L}_{\text{blk}}$ is up-weighted to improve the blocked-class F1-score. The weights $\lambda_{\text{beam}}$, $\lambda_{\text{blk}}$, and $\lambda_{\text{pose}}$ are selected so that the individual loss terms have comparable magnitude while prioritizing beam alignment. A ReduceLROnPlateau scheduler monitors validation loss, and the checkpoint with the lowest validation loss is used for all test results in Section~\ref{sec:results}.
\subsection{LiDAR Map and Trajectory Visualization}
To complement numerical pose metrics, we construct an offline LiDAR map and overlay the predicted trajectory. LiDAR scans are accumulated in the BS-centred local frame using the dataset-provided synchronization and calibration (and, when applicable, reference position information). The reference trajectory $\{\mathbf{s}_t\}$ and predicted trajectory $\{\hat{\mathbf{s}}_t\}$ are then plotted on a top-down projection of the aggregated point cloud. Although our model regresses only planar coordinates and does not perform scan-to-map optimization, this visualization provides a SLAM-style sanity check on geometric consistency and exposes systematic bias between $\hat{\mathbf{s}}_t$ and $\mathbf{s}_t$.
\section{Results and Discussion}
\label{sec:results}
We evaluate on DeepSense~6G Scenario~31 (7012 synchronized snapshots), split by sequence into training/validation/test sets (approximately 70\%/15\%/15\%) to reduce temporal leakage. All models are trained for 40 epochs with the multi-task loss in~\eqref{eq:learn}, and the best validation checkpoint is used for testing.
\subsection{Training Behaviour}
\begin{figure}[htbp]
  \centering
\includegraphics[width=0.7\columnwidth]{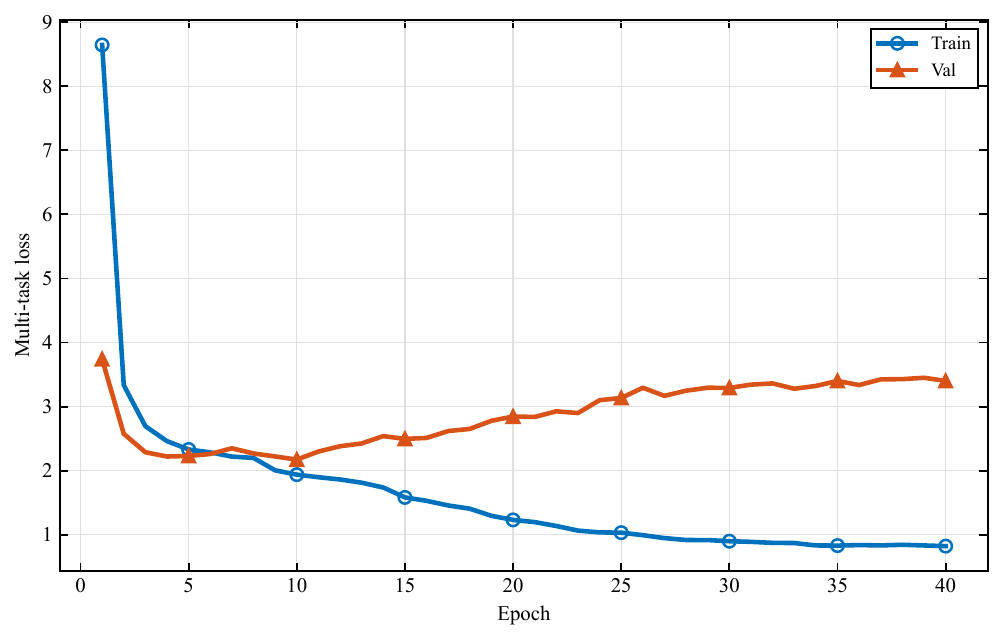}
  \caption{Training and validation multi-task loss over epochs for the multimodal model.}
  \label{fig:loss_curve}
\end{figure}

Fig.~\ref{fig:loss_curve} shows the joint training and validation loss over 40 epochs. Both curves decreases rapidly during the first 10 epochs after which validation curves increases, but the training curve continue to decrease, indicating effective learning without severe overfitting.
\subsection{Beam Alignment and Spectral Efficiency}
\begin{figure}[htbp]
  \centering
\includegraphics[width=0.7\columnwidth]{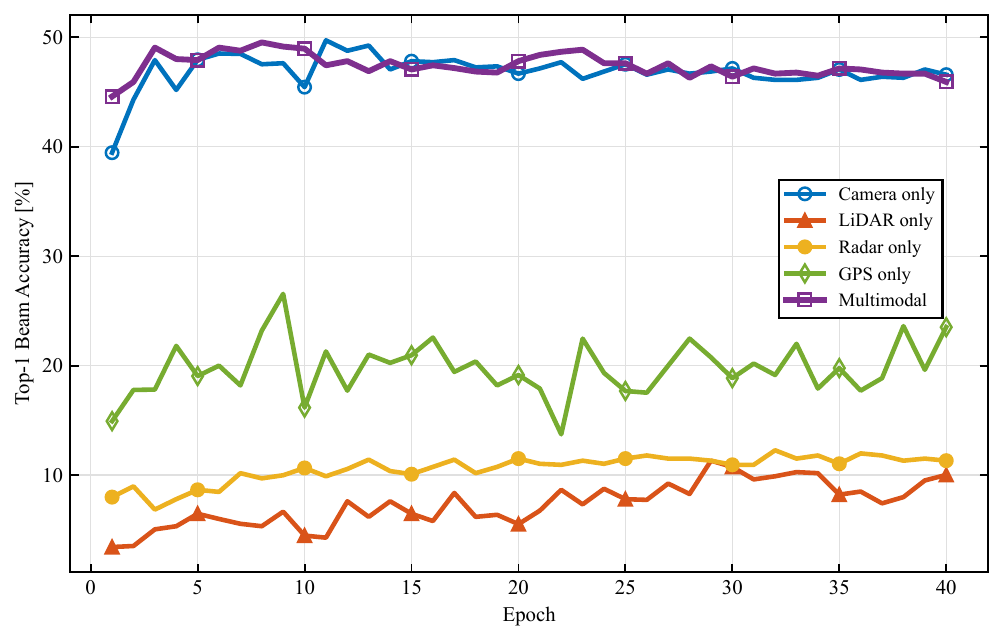}
  \caption{Validation Top-1 beam accuracy versus epoch for the multimodal model and unimodal baselines.}
  \label{fig:beam_top1_all}
\end{figure}

\begin{figure}[htbp]
  \centering
\includegraphics[width=0.7\columnwidth]{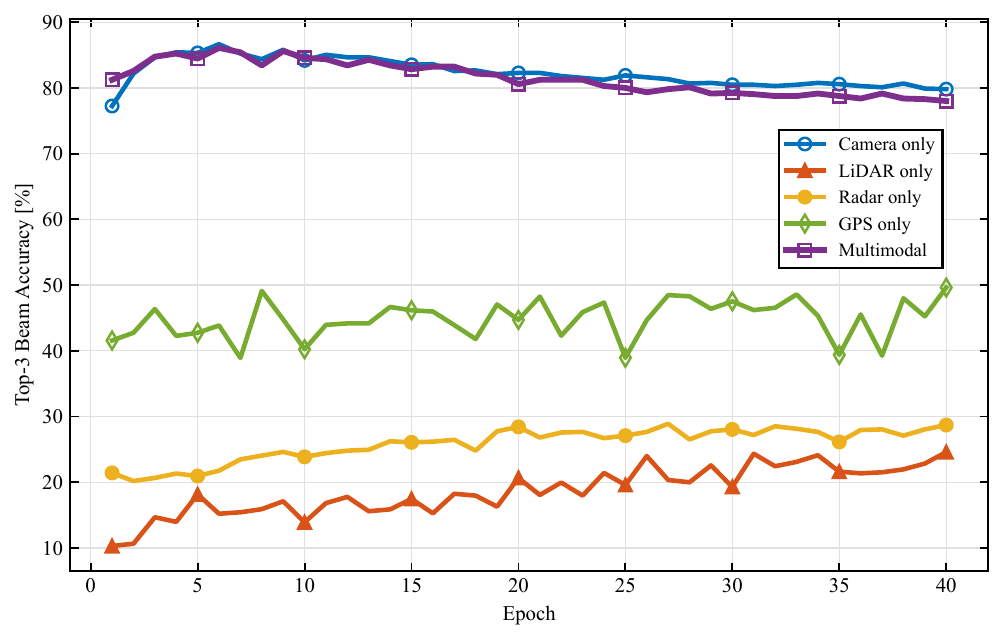}
  \caption{Validation Top-3 beam accuracy versus epoch for the multimodal model and baselines.}
  \label{fig:beam_top3_all}
\end{figure}

\begin{figure}[htbp]
  \centering
\includegraphics[width=0.7\columnwidth]{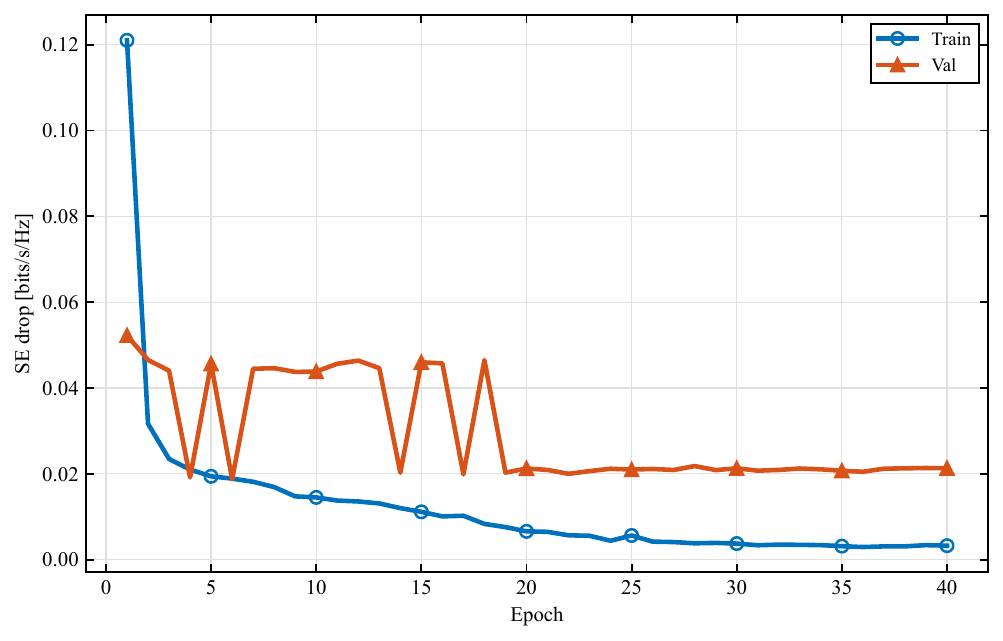}
  \caption{Average SE drop versus epoch for training and validation splits.}
  \label{fig:se_drop_curve}
\end{figure}

Fig.~\ref{fig:beam_top1_all} and \ref{fig:beam_top3_all} compare Top-1 and Top-3 beam accuracy across models on the validation set. On the held-out test set, the proposed multimodal network achieves 50.92\% Top-1 and 86.50\% Top-3 accuracy, marginally outperforming the best unimodal alternative (camera-only: 50.79\% / 86.03\%).
The spectral-efficiency impact of residual misalignment is captured by the SE-drop metric in~\eqref{eq:se_drop_def}. Fig.~\ref{fig:se_drop_curve} shows that the validation SE drop decreases throughout training. On the test set, the multimodal model attains an average SE drop of 0.018\,bits/s/Hz, slightly better than the camera-only baseline (0.019\,bits/s/Hz).
The spectral-efficiency impact of residual misalignment is captured by the SE-drop metric in~\eqref{eq:se_drop_def}. Fig.~\ref{fig:se_drop_curve} shows that the validation SE drop decreases throughout training. On the test set, the multimodal model attains an average SE drop of 0.018\,bits/s/Hz, slightly better than the camera-only baseline (0.019\,bits/s/Hz).
\subsection{Blockage Detection}
\begin{figure}[htbp]
  \centering
\includegraphics[width=0.7\columnwidth]{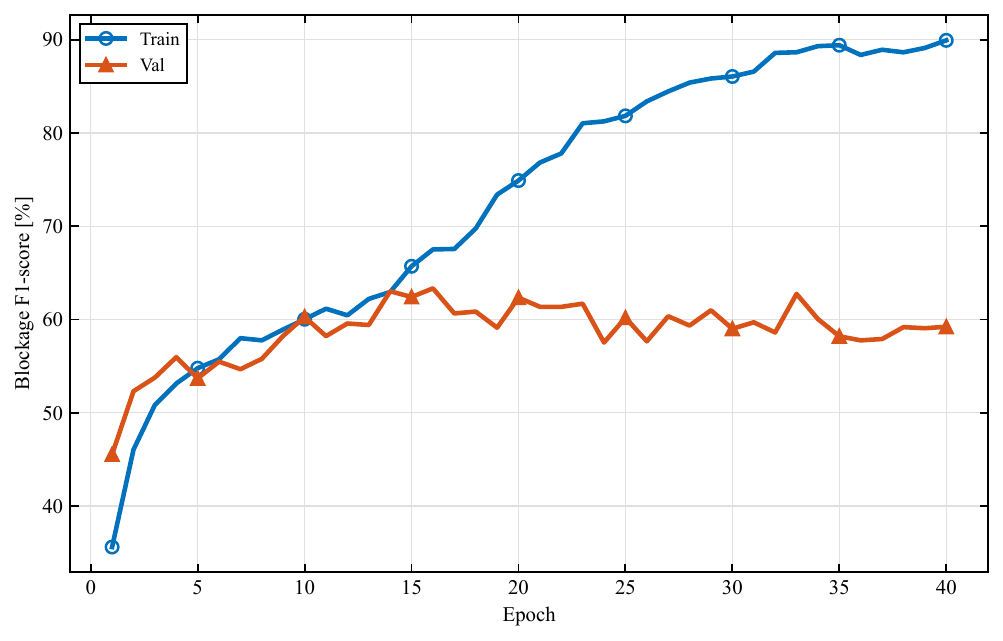}
  \caption{Blockage F1-score versus epoch for training and validation splits.}
  \label{fig:blockage_f1_curve}
\end{figure}

Blockage detection is measured by accuracy and blocked-class F1-score; because blocked samples are relatively scarce in Scenario~31, F1 is the most informative metric. As shown in Fig.~\ref{fig:blockage_f1_curve}, the multimodal model's validation F1 increases steadily during training. On the test set, it achieves 63.35\% blocked-class F1, improving over the camera-only baseline (59.04\%) and all other unimodal models.
\subsection{Localization and SLAM-Style Behaviour}
\begin{figure}[htbp]
  \centering
\includegraphics[width=0.6\columnwidth]{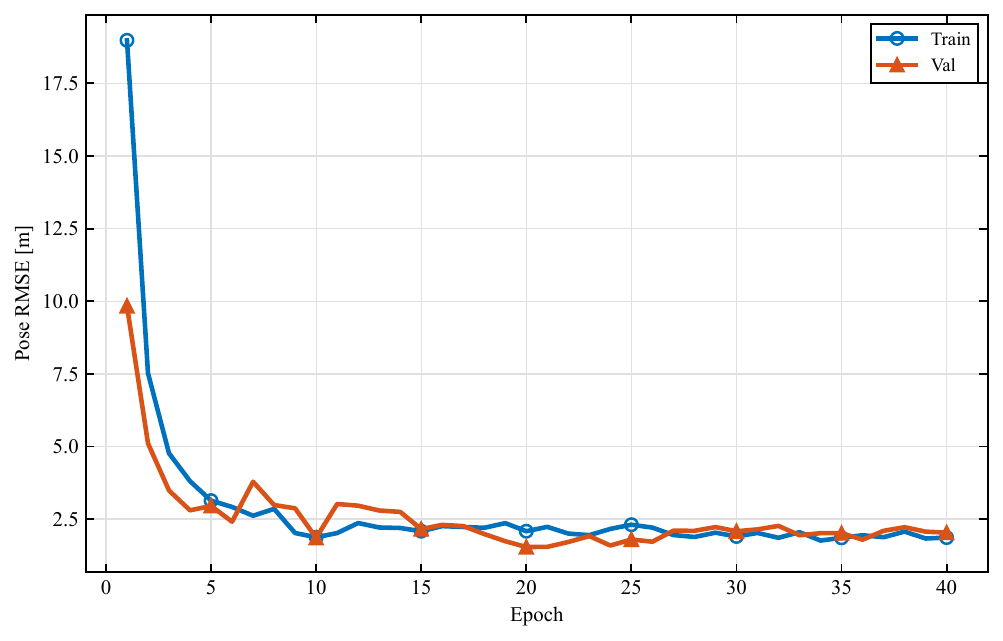}
  \caption{2D position RMSE versus epoch for training and validation splits.}
  \label{fig:pose_rmse_curve}
\end{figure}

\begin{figure}[htbp]
  \centering
\includegraphics[width=0.6\columnwidth]{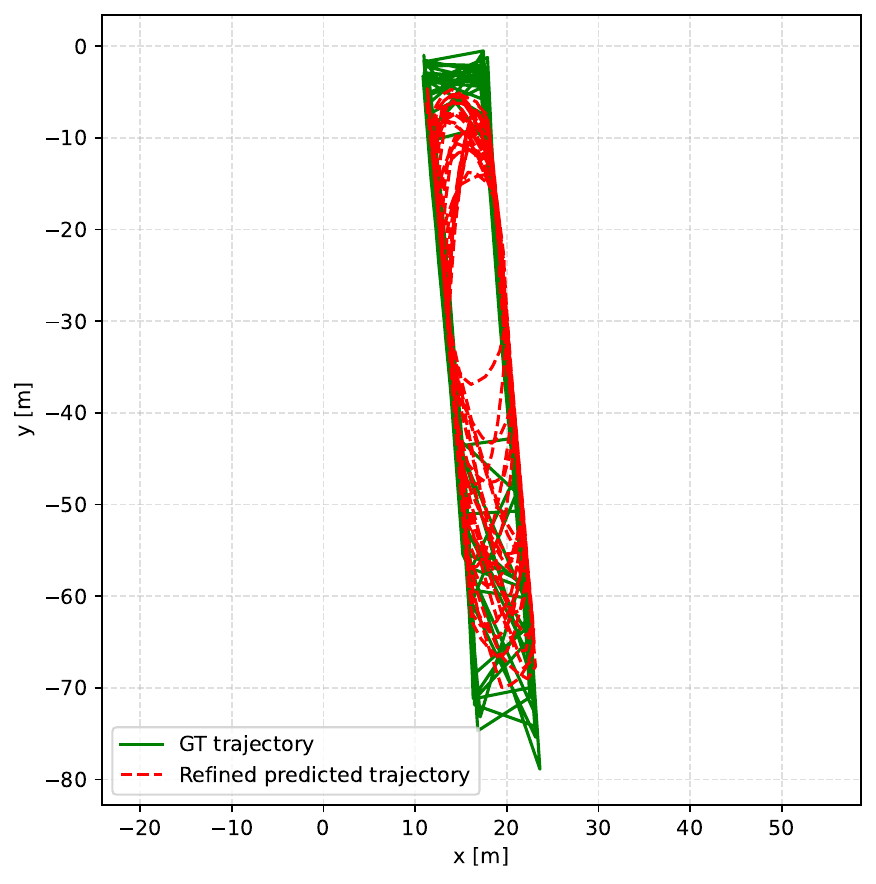}
  \caption{Top-down LiDAR map with overlaid reference trajectory and predicted trajectory.}
  \label{fig:lidar_traj_overlay}
\end{figure}

The pose head is evaluated using the 2D RMSE in~\eqref{eq:pose_rmse_def} between predicted positions and the reference trajectory. Fig.~\ref{fig:pose_rmse_curve} shows that both training and validation RMSE decrease and stabilize below 1.5\,m. On the test set, the multimodal model attains 1.33\,m RMSE, improving over the camera-only model (2.10\,m) and all other unimodal baselines.
Fig.~\ref{fig:lidar_traj_overlay} provides a qualitative SLAM-style visualization by overlaying the predicted and reference trajectories on the aggregated LiDAR map. The predicted trajectory follows the dominant street corridor captured by the map and exhibits limited drift over the drive, supporting the geometric consistency of the learned representation.
\subsection{Comparison with Unimodal Baselines}
Table~\ref{tab:test_metrics} summarizes test-set performance. Vision is the strongest single modality for beam management in this scenario, but multimodal fusion provides the best overall trade-off: it matches (and slightly improves) camera-only beam accuracy and SE drop while delivering larger gains in blockage awareness (63.35\% vs.\ 59.04\% F1) and localization (1.33\,m vs.\ 2.10\,m RMSE).
\begin{table}[t]
  \centering
  \caption{Test performance on DeepSense~6G Scenario~31.}
  \label{tab:test_metrics}
  \renewcommand{\arraystretch}{0.8}
  \setlength{\tabcolsep}{3.5pt}
  \footnotesize
  \begin{tabular}{lccccc}
    \toprule
    Model & Top1[\%] & Top3[\%] & $\overline{\Delta\mathrm{SE}}$ & F1$_\mathrm{blk}$[\%] & RMSE[m] \\
    \midrule
    mmWave   &  6.37 & 17.59 & 0.371 &  0.00 & 18.28 \\
    LiDAR    &  9.79 & 20.82 & 0.288 & 36.26 & 16.52 \\
    Radar    & 14.96 & 28.74 & 0.264 & 32.79 & 15.49 \\
    GPS      & 20.53 & 44.65 & 0.208 & 48.15 &  4.49 \\
    Camera   & 50.79 & 86.03 & 0.019 & 59.04 &  2.10 \\
    \midrule
    \textbf{Multi. (prop.)} & \textbf{50.92} & \textbf{86.50} & \textbf{0.018} & \textbf{63.35} & \textbf{1.33} \\
    \bottomrule
  \end{tabular}
\end{table}
Overall, the multimodal model is the strongest performer across tasks. The small Top-1/Top-3 beam-accuracy gap between multimodal and camera-only suggests that vision captures most geometric cues for beam selection in this scenario, while fusing LiDAR, radar, GNSS, and radio history improves robustness, particularly for blockage detection and pose estimation.
\section{Conclusion}
We presented an occlusion-aware multimodal learning framework, inspired by SLAM concepts, for beam-aligned mmWave V2I on the real 60\,GHz DeepSense~6G Scenario~31 dataset. A Transformer-based fusion network integrates RGB images, LiDAR point clouds, radar features, GNSS, and short-term mmWave power history to jointly predict the receive beam index, blockage state, and 2D vehicle position. Experimental results confirm that multimodal fusion provides the strongest overall performance, achieving 50.92\% Top-1 and 86.50\% Top-3 beam accuracy with only 0.018\,bits/s/Hz average spectral-efficiency loss, a 63.35\% blocked-class F1-score, and a 1.33\,m localization RMSE. While camera-only is a competitive baseline for beam selection, additional modalities improve robustness, particularly for blockage awareness and localization. Future work will study GNSS-free localization, stronger temporal modeling for high-mobility scenes, and closed-loop integration with practical beam management and handover procedures for real-time vehicular deployments.
\section*{Acknowledgment}
This work was jointly supported by the African Center of Excellence in Internet of Things (ACEIoT) University of Rwanda,
Regional Scholarship and Innovation Fund (RSIF), and National Research Foundation of Korea under Grant RS-2024-00409492.
\bibliographystyle{IEEEtran}
\bibliography{IEEEabrv,References}
\end{document}